# CheckSupport: A Local LLM-Powered Tool for Automated Manuscript Submission Checklist Selection and Completion

Satvik Tripathi[1,#], Don Enwerem[2], Kevin Song[3], Kristian Quevada[4], Jacinta Arnold[5], Tessa S. Cook[1]

1. Department of Radiology, Perelman School of Medicine at University of Pennsylvania, Philadelphia, PA, USA
2. Department of Computer Science, Drexel University, Philadelphia, PA, USA
3. Department of Computer and Information Science, School of Engineering and Applied Science, University of Pennsylvania, Philadelphia, PA 19104
4. Department of Radiology, Cooper University Hospital, Camden, NJ, USA
5. University of California Davis Graduate School of Management, Davis, CA

#Satvik.tripathi@pennmedicine.upenn.edu
3400 Spruce St, Philadelphia, PA, 19104

**Abstract**

Transparent and standardized reporting is essential for reproducible scientific research, yet adherence to reporting guidelines remains inconsistent because of the manual effort required to select and complete checklists. We present CheckSupport, an open-source, locally deployable system that uses large language models to automate the recommendation of reporting checklists and the evidence-grounded completion of checklists for scientific manuscripts. CheckSupport employs a staged prompting strategy that decomposes reporting workflows into constrained inference tasks, prioritizing faithful extraction over generative text synthesis. All inference is performed locally using instruction-tuned models, preserving data privacy and enabling reproducible, auditable workflows. Evaluated on a corpus of peer-reviewed manuscripts, CheckSupport achieved 90% overall accuracy for checklist recommendations and 88% overall accuracy for item-level completion while operating on CPU-only hardware. On average, the wall-clock time per manuscript was 12.5 seconds, including the checklist recommendation and full checklist completion. These results demonstrate that large language models, when applied as structured inference components, can reduce reporting burden and support more transparent and reproducible scientific reporting across disciplines.

**Github:** https://github.com/Penn-RAIL/checkSupport

## 1. Introduction

Transparent and standardized reporting is fundamental to the credibility, reproducibility, and reuse of scientific research. Reporting guidelines such as CONSORT, DEAL, PRISMA, STARD, and other domain-specific checklists were developed to ensure that essential methodological details are communicated clearly and consistently[1,2,3]. When followed, these frameworks improve study interpretability, facilitate peer review, and support evidence synthesis[4,5,6].

Despite their widespread endorsement, reporting guidelines are often not fully implemented in practice. Identifying the appropriate checklist for a given study and completing it accurately requires substantial manual effort and domain expertise. As scientific studies increasingly incorporate complex computational methods, multi-stage pipelines, and large datasets, the burden of comprehensive reporting continues to grow for both authors and reviewers.

Recent advances in large language models (LLMs) have generated interest in automating components of the scientific writing and review process[7,8]. Prior work has explored the use of LLMs for manuscript summarization, language editing, and reviewer assistance[9,10,11]. However, structured reporting support remains relatively underexplored. Existing tools are largely manual or rely on cloud-based LLM services, raising concerns about data privacy, reproducibility, cost, and long-term accessibility[12,13,14].

To address these challenges, we introduce CheckSupport. This open-source command-line tool leverages locally deployed large language models to automate two critical steps in reporting guideline workflows: (1) recommending the most appropriate reporting checklist for a manuscript and (2) completing the selected checklist using content grounded directly in the manuscript text. By operating entirely offline and using structured prompting aligned with Equator Network criteria, CheckSupport aims to reduce the burden on authors and reviewers while supporting transparent, reproducible reporting across scientific disciplines.

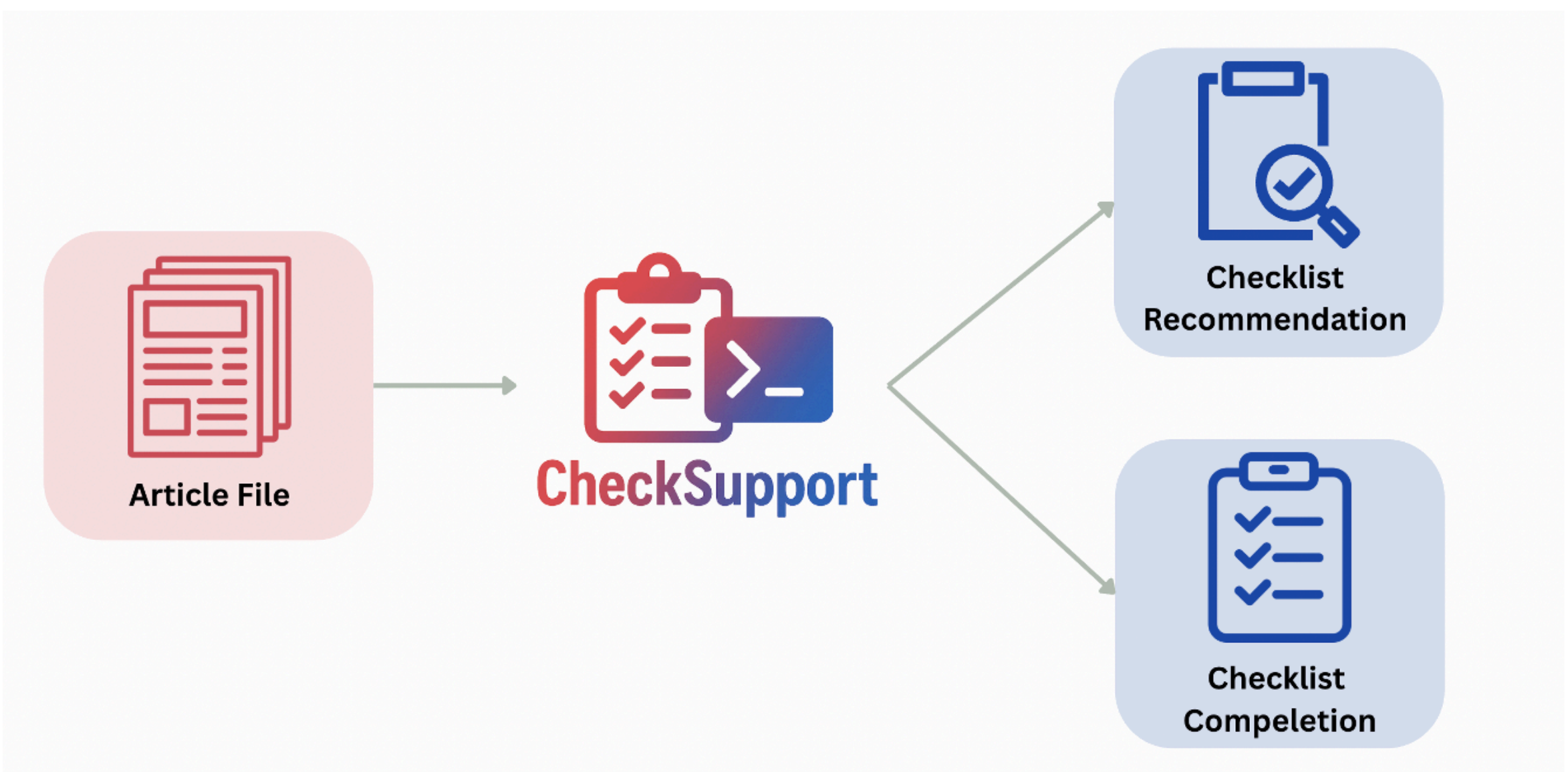

## 2. System Overview

### 2.1 System Architecture

CheckSupport is an open-source software system designed to automate the selection and completion of reporting checklists for scientific manuscripts. The system is implemented as a Python-based command-line interface and is intended to integrate into existing research authoring and editorial workflows. All components of the system operate locally, without reliance on cloud-hosted services or external application programming interfaces.

The overall architecture follows a modular pipeline consisting of manuscript ingestion, checklist identification, checklist parsing, content extraction, and report generation. Each stage is implemented as an independent module, enabling extensibility and facilitating future adaptation to additional reporting standards or model backends.

### 2.2 Design Principles

The design of CheckSupport is guided by three primary principles: privacy preservation through fully local execution, reproducibility through deterministic, transparent processing, and extensibility through a modular system design. These principles enable CheckSupport to serve as a general-purpose framework for structured reporting across scientific disciplines.

### 2.3 Input and Output Interfaces

CheckSupport accepts manuscript files in PDF, DOCX, or plain text formats. Input documents undergo format-specific text extraction and syntax normalization before analysis. Reporting checklists may similarly be provided as PDF, DOCX, or text files, allowing the system to support both standardized guidelines and custom checklist definitions.

The system produces two primary outputs. For the checklist recommendation, the output is the name of the most appropriate reporting guideline. For checklist completion, the output is a structured, fully populated checklist rendered as a formatted PDF document suitable for submission or peer review.

### 2.4 Core Functional Workflows

The system supports two principal workflows that may be executed independently or in sequence. The first workflow analyzes manuscript content and recommends an appropriate reporting checklist. The second workflow populates a selected checklist by extracting and synthesizing relevant information from the manuscript text. Within the checklist completion workflow, checklist documents are parsed into sections and individual items. These elements serve as the structural scaffold for downstream processing, enabling section- and item-level analysis during checklist population.

### 2.5 Local Model Execution

All language-model inference is performed locally using the Ollama framework, which enables the execution of open-source large language models on consumer-level hardware. By default, CheckSupport uses an instruction-tuned model, though users may specify alternative models at runtime.

Local execution ensures that manuscript content remains on the user's system, addressing data privacy concerns and enabling reproducible and auditable workflows. Inference parameters are explicitly controlled to promote consistency and reduce variability in model outputs.

### 2.6 Checklist Representation and Parsing

CheckSupport includes built-in templates for widely used reporting guidelines and supports user-provided checklists. Checklist documents are parsed programmatically to identify sections, item text, and information-driven guidance, where available. Both structured templates and loosely formatted checklist documents are supported. Parsed checklist elements are represented internally as hierarchical data structures, which are subsequently used to guide targeted extraction and checklist completion.

## 3. Prompting and Inference Strategy

CheckSupport implements a staged prompting-and-inference framework that decomposes the selection and completion of reporting checklists into a sequence of constrained language-model invocations. The system is designed to treat large language models (LLMs) as controlled inference components rather than free-form generators, with explicit limits on context, output structure, and inference parameters at each stage. This approach prioritizes faithful extraction, reproducibility, and auditability over generative fluency.

### 3.1 Overview of the Inference Pipeline

The inference pipeline consists of four sequential stages: (1) checklist recommendation, (2) checklist preprocessing and guidance generation, (3) section-level text extraction, and (4) item-level answer generation. Each stage is implemented as an independent model call with task-specific prompt templates, temperature settings, and context constraints. Outputs from earlier stages are explicitly passed forward to condition subsequent prompts, enabling progressive narrowing of the task scope and context.

By separating classification, interpretation, and extraction into discrete steps, the system avoids reliance on long-context prompts and reduces error propagation across tasks. All inference is performed locally using instruction-tuned language models via the Ollama framework.

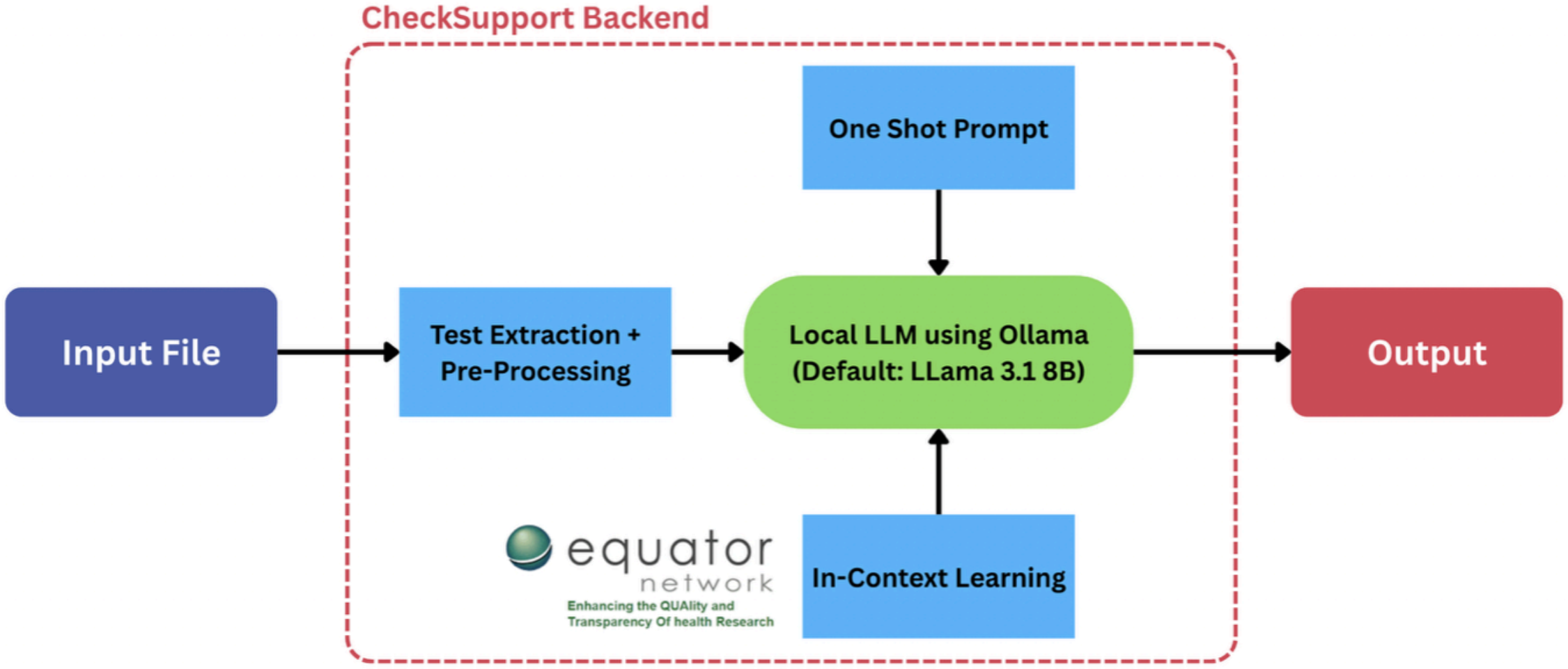

### 3.2 Language Models and Runtime Configuration

CheckSupport is model-agnostic by design and supports any instruction-tuned model compatible with Ollama. In this study, the default model used was Llama 3.1, selected for its strong performance on structured reasoning tasks, modest parameter size, and suitability for execution on commodity hardware. Additional supported models include Gemma, Mistral, and any other Ollama model, which can be specified at runtime via command-line arguments.

All model invocations are performed synchronously via the Ollama generation API, with streaming disabled, to ensure complete, deterministic responses. Inference parameters are explicitly controlled at each stage of the pipeline. Low temperatures are used for classification and item-level extraction tasks to reduce variance, whereas higher temperatures are reserved for intermediate guidance-generation steps that benefit from broader semantic exploration.

Manuscript text is truncated and processed in fixed-size windows to accommodate varying model context limits. This design avoids dependence on extended-context models and ensures consistent behavior across different hardware and model configurations.

### 3.3 Checklist Recommendation as Constrained Classification

Checklist recommendation is formulated as a one-shot memory, single-label classification task. The model is provided with a truncated excerpt of the manuscript text, limited to the first 2,000 characters, along with a fixed list of candidate reporting checklists derived from available templates. The prompt explicitly instructs the model to select exactly one checklist name from the provided list and to return only that checklist name. Restricting both the input length and output format serves two purposes. First, it focuses the model on high-level study characteristics such as design and objectives. Second, it simplifies downstream parsing and prevents ambiguous or multi-label responses. Inference at this stage is performed with a low temperature (0.2) to encourage deterministic behavior.

### 3.4 Checklist Preprocessing and Guidance Generation

Once a checklist is selected and parsed into sections and items, CheckSupport performs a preprocessing step to generate extraction guidance. This step uses the language model to infer which information should be sought for each checklist section and how it is typically expressed in scientific manuscripts. Guidance generation is performed in two passes. In the first pass, the model is prompted with the complete list of section names and asked to provide general guidance on the types of information associated with each section. In the second pass, section-specific prompts request more detailed advice, including typical manuscript locations and indicative phrases or concepts. These prompts are administered at a higher temperature (0.7) to elicit comprehensive and descriptive responses. The generated guidance is stored internally and used to condition subsequent extraction prompts. It is not presented directly to the user and does not constitute checklist content.

### 3.5 Section-Level Text Extraction

For each checklist section, the system prompts the model to extract the most relevant portions of the manuscript text. The prompt includes the section name, the section-specific guidance generated during preprocessing, and a bounded excerpt of the manuscript. Manuscript text is processed in fixed windows of 5,000 characters to avoid exceeding model context limits. If the extracted output falls below a minimum length threshold, the system automatically retries extraction using subsequent text windows. Inference at this stage uses a moderate temperature (0.3) to balance recall and precision. The resulting section-level extracts are treated as canonical context for downstream item-level completion and are reused for all items within a section.

### 3.6 Item-Level Answer Generation

Checklist completion is executed at the granularity of individual checklist items. For each item, the model is prompted with (1) the checklist item text, (2) any item-specific instructions parsed from the source checklist, (3) section-level extracted manuscript content, and (4) a constrained window of additional contextual text (≤1,000 characters). Prompts are explicitly structured to elicit concise, evidence-grounded responses restricted to the provided inputs, with instructions to explicitly indicate when required information is absent. Inference is performed under controlled decoding settings (temperature = 0.5) with predefined stop sequences to suppress verbosity and prevent uncontrolled continuation beyond the target response span. Each checklist item is processed independently, eliminating inter-item dependency and mitigating error propagation. This modular design enables deterministic behavior, improves robustness, and supports fine-grained auditing and traceability at the level of individual checklist elements.

### 3.7 Output Control and Hallucination Mitigation

Multiple safeguards are implemented to minimize hallucination and enforce faithful, text-grounded extraction. Context windows are strictly bounded at each inference stage, and prompts consistently reinforce reliance on the provided manuscript content. The generation of explicit null responses is permitted and encouraged when required information is absent, preventing speculative completion. Decoding parameters are stage-specific, with temperature tuned to task demands, and predefined stop sequences are applied to constrain output length and terminate generation at the appropriate boundary. By decomposing checklist completion into discrete, independently executable inference steps and enforcing strict prompt-level constraints, CheckSupport prioritizes factual accuracy, interpretability, and traceability over generative fluency or narrative synthesis.

## 4. Evaluation

### 4.1 Manuscript Corpus

CheckSupport was evaluated on a corpus of peer-reviewed scientific manuscripts representing a range of study designs and reporting requirements. The corpus was derived from a PubMed-indexed search for the term “AI in radiology,” limited to publications from 2025, and the top 100 manuscripts were randomly selected. Manuscripts were selected to ensure coverage across reporting checklist categories, including studies for which a structured reporting guideline was applicable and manuscripts for which no standard checklist applied. All manuscripts were provided to the system in their original submission formats (PDF or DOCX) without manual preprocessing beyond text extraction.

To establish reference standards, each manuscript was independently reviewed by domain experts with more than 5 years of research experience, who were assigned a ground-truth reporting checklist category. For manuscripts deemed checklist-eligible, experts also manually completed the corresponding checklist to serve as a benchmark for automated checklist completion.

### 4.2 Ground-Truth Annotation

Checklist eligibility and checklist item content were annotated by expert reviewers familiar with reporting guideline requirements. For the checklist recommendation, manuscripts were assigned to one of three categories: checklist type A, checklist type B, or not applicable. Checklist type A encompassed manuscripts involving the development of new LLM structures, advanced techniques, or fine-tuning large datasets. Checklist type B encompassed manuscripts with more practical LLM applications, such as content generation or classification, with minimal customization. The ‘not applicable’ category captured manuscripts that do not meet the criteria of checklists A or B. For checklist completion, expert annotations consisted of item-level responses indicating whether required information was present and, if so, the corresponding manuscript content. These annotations served as the reference standard for both checklist selection accuracy and checklist completion accuracy.

### 4.3 Evaluation Tasks

Evaluation focused on two primary tasks corresponding to the core system functionalities. The first task assessed the checklist recommendation performance. For each manuscript, the checklist suggested by CheckSupport was compared against the expert-assigned reference category. Performance was quantified using overall accuracy and category-specific accuracy. The second task evaluated checklist completion performance. For manuscripts assigned an applicable checklist, automated item-level responses generated by CheckSupport were compared against expert-completed checklists. An item was considered correct if the automated response accurately reflected the presence or absence of required information and, when present, correctly summarized the corresponding manuscript content.

### 4.4 Performance Metrics

Checklist recommendation performance was measured using accuracy at the manuscript level. Checklist completion performance was measured using item-level accuracy, defined as the proportion of checklist items for which the system-generated response matched the expert annotation.

In addition to accuracy, failure modes were qualitatively analyzed to characterize common sources of error, including ambiguous manuscript descriptions, incomplete reporting, and extraction errors.

### 4.5 Runtime and Hardware Configuration

All evaluations were conducted using locally hosted language models executed via the Ollama framework. Inference was performed on CPU-only hardware to reflect realistic usage scenarios in typical research and editorial environments. No GPU acceleration was used. Runtime performance was measured as end-to-end wall-clock time per manuscript for both checklist recommendation and checklist completion workflows. Model configuration, temperature settings, and context window sizes were fixed across all evaluations to ensure consistency.

## 5. Results

### 5.1 Checklist Recommendation Performance

CheckSupport demonstrated strong performance in recommending appropriate reporting checklists across the corpus of evaluated manuscripts. Overall, checklist recommendation accuracy was 90.0%, indicating that the system correctly identified the expert-assigned checklist category for the majority of manuscripts. Performance varied modestly by category. For manuscripts assigned to checklist category A, the system achieved 86.7% accuracy. For checklist category B manuscripts, accuracy was 100%, reflecting reliable classification in studies with clearly defined reporting requirements. For manuscripts deemed not applicable to a standard reporting checklist, accuracy was 80.0%. Most misclassifications occurred in borderline cases with limited methodological detail or hybrid study designs.

### 5.2 Checklist Completion Accuracy

Among manuscripts eligible for checklist completion, CheckSupport achieved an overall item-level accuracy of 88.0%. Performance differed across checklist categories, with 80.0% accuracy for category A manuscripts and 93.3% accuracy for category B manuscripts. Incorrect responses were most commonly associated with checklist items requiring nuanced interpretation of implicit methodological details or information distributed across multiple manuscript sections. In contrast, checklist items tied to explicitly reported methodological elements were completed with high fidelity.

### 5.3 Error Analysis

Qualitative review of incorrect outputs revealed three primary error modes. First, ambiguity in the manuscript language led to partial or conservative extraction when relevant information was present but stated indirectly. Second, incomplete reporting in the source manuscript led to null responses, which the system correctly flagged but were counted as errors under strict accuracy criteria. Third, occasional extraction failures occurred when relevant information fell outside the bounded context windows used during section-level extraction.

Notably, no instances of unsupported or fabricated content were observed, consistent with the system's emphasis on grounded extraction and explicit null responses.

### 5.4 Runtime Performance

End-to-end runtime performance was evaluated on CPU-only hardware. Average wall-clock time per manuscript was 12.5 seconds, including checklist recommendation and full checklist completion. Runtime scaled approximately linearly with manuscript length and checklist size, with section-level extraction accounting for most of the inference time. These results indicate that CheckSupport can be used in routine authoring and editorial workflows without requiring specialized computing resources.

## 6. Discussion

This work illustrates how large language models can be integrated into scientific workflows in a way that prioritizes transparency, accountability, and reproducibility over the generation of primary content. As reporting requirements continue to expand in response to increasingly complex research methodologies, the burden placed on authors, reviewers, and editors has grown substantially. Tools like CheckSupport offer a complementary approach that shifts part of this burden from manual compliance toward structured, assistive automation.

More broadly, this study highlights an emerging role for language models as infrastructure for research governance rather than content creation. By focusing on checklist selection and evidence-grounded extraction, CheckSupport reframes the use of LLMs from narrative generation to methodological support. This distinction is critical in scientific contexts, where the primary objective is not eloquence but faithful representation of what was actually done and reported. The system's conservative handling of missing information underscores this principle by making gaps in reporting explicit rather than implicitly filling them.

The use of fully local model execution further reflects a growing recognition that privacy, reproducibility, and long-term sustainability are central considerations for AI-enabled research tools. As concerns around data leakage, proprietary dependencies, and opaque inference pipelines increase, locally deployable systems provide a path toward greater institutional trust and broader adoption. The ability to audit intermediate steps and reproduce outputs under fixed configurations aligns closely with existing norms of scientific scrutiny.

From a field-wide perspective, automated reporting support can influence not only author behavior but also editorial and peer-review practices. Lowering the barrier to checklist completion may improve adherence to reporting standards, reduce reviewer fatigue, and support more consistent evaluation of methodological rigor. Over time, such tools could contribute to a cultural shift in which reporting guidelines are viewed less as administrative obligations and more as integral components of the research process.

At the same time, this work reinforces the importance of human oversight. Automated systems cannot resolve ambiguity inherent in poorly reported studies, nor should they substitute for expert judgment. Instead, their value lies in making structure explicit, surfacing omissions, and enabling researchers and reviewers to focus their attention where it is most needed. As language models continue to evolve, the most impactful applications in science may be those that enhance rigor and accountability rather than replace human interpretation.

**7. Conclusion**

CheckSupport demonstrates how large language models can support scientific reporting in a structured, accountable manner. By framing checklist selection and completion as constrained inference tasks and executing all inference locally, the system prioritizes transparency, reproducibility, and privacy over generative flexibility. This work illustrates a broader role for language models as infrastructure for research governance rather than content creation. As reporting requirements grow in complexity, tools like CheckSupport can reduce compliance burden while preserving human oversight, contributing to more consistent and reproducible scientific reporting across disciplines.